\documentclass[aps,pre,twocolumn]{revtex4-1}
\usepackage{amssymb}
\usepackage{amsmath,bm}
\usepackage{graphicx,color}
\usepackage{bbm}
\usepackage[bottom]{footmisc}
\usepackage{multirow}
\newcommand{\B}[1]{{\bm{#1}}}

\begin{document}
\title{Scaling Theory for W\"ohler plots in amorphous solids under cyclic forcing}
\author{Bhanu Prasad Bhowmik $^1$, 
	H.G.E.Hentschel$^{1,2}$ and Itamar Procaccia$^{1,3}$}
\affiliation{$^1$Dept. of Chemical Physics, The Weizmann Institute of
	Science, Rehovot 76100, Israel\\
$^2$ Dept. of Physics, Emory University, Atlanta Ga. 30322, USA\\$^3$  Center for OPTical IMagery Analysis and Learning, Northwestern Polytechnical University, Xi'an, 710072 China. }

\begin{abstract}
In mechanical engineering W\"ohler plots serve to measure the average number of load cycles before materials break, as a function of the maximal stress in each cycle. Although such plots are prevalent in engineering for more than 150 years, their theoretical understanding is lacking. Recently a scaling theory of W\"ohler plots in the context of cyclic {\em bending} was offered \cite{21BHP}. Here we elaborate further on cyclic bending and extend the considerations to cyclic {\em tensile} loads on an amorphous strip of material;   the scaling theory applies to both types of cyclic loading equally well. On the basis of atomistic simulations we conclude that the crucial quantities to focus on are the {\em accumulated damage} and the {\em average damage per cycle}. The dependence of these quantities on the loading determines the statistics of the number of cycles to failure. Finally we consider the probability distribution functions of the number of cycles to failure and demonstrate that the scaling theory allows prediction of these distributions at one value of the forcing amplitude from measurements and another value. 
\end{abstract}
\maketitle

\section{Introduction} 
One of the better studied phenomena in mechanical engineering is the so called ``fatigue" that results from oscillatory applied strains on devices and materials \cite{54Mil,66Dow,89DBS,94KC}. A standard measure of the resilience to such oscillatory loading are the so-called ``W\"ohler diagrams" or ``s-n plots" (shorthand for ``stress vs number of cycles"), which display the exponential decrease of the number of cycles before failure upon increasing the maximal external load.  This exponential dependence was first discovered by of W\"ohler who investigated the famous 1842 train crash in Versailles, France.  Surprisingly, in spite of the long time elapsing since and a variety of attempts to provide theoretical explanations to W\"ohler diagrams, no accepted theory that encompasses different types of cyclic loading has emerged. To some extent this may be attributed to the tendency of engineers to look for ``mechanisms". Indeed,  in an early paper Freudenthal, Gumbel and Gough \cite{53FGG} criticized this tendency, writing: ``Practically all existing fatigue theories ... operate on the assumption that fatigue can be explained in terms of a single mechanism. The fact is not considered that one mechanism alone can hardly be expected to describe a phenomenon that is the result of force- and time-dependent processes on the microscopic and submicroscopic level, which are associated with the existence of highly localized textural stress fields, defects and anomalies in the ideal structure of the material. The usual engineering abstraction of such a material as a continuous, homogeneous, isotropic, elastic body therefore precludes any effective theoretical approach to fatigue."  In our work we indeed follow up on these comments. In a recent Letter \cite{21BHP} we offered a scaling theory that avoids the specific mechanistic approach in favor of identifying measurable quantities of damage
incurred to the material. We employed atomistic simulations of a strip of athermal amorphous solid to focus on ``progressive damage" \cite{87CC,07LH,16TV}, on its statistics and on its dependence on the amplitude of the load. This allowed us to offer a scaling theory for the dependence of the mean number of cycles to failure, as a function of the load amplitude (W\"ohler diagrams). 

In this paper we elaborate further on this approach, where instead of focusing on detailed mechanistic studies, we use scaling concepts to provide a predictive theory based on a minimal number of material properties. This generic approach allows us to deal with both bending and tensile loading in precisely the same way, underlining the generality of the approach and its independence of details mechanisms. A definite advantage of the scaling approach will be the ability to predict the statistics
of failure at varying amplitudes of loading from the measurement of one 
value. This is demonstrated below. 

The structure of the paper is as follows: in Sect.~\ref{simulations} we describe the system preparation and the cyclic bending and tensile protocols. Section~\ref{damage} provides a discussion
of the measured quantities that control the progress of fatigue, i.e. the damage per cycle
and the accumulated damage. In Sect.~\ref{exponents} we present the data for both bending and tensile protocols, and identify the quantities that theory should relate to each other. That is done in Sect.~\ref{scaling} that shows how to estimate the parameters appearing in W\"ohler plots from measurements of the
relevant length scales and appropriate measures of damage. Section~\ref{statistics} deals with the probability distribution functions (PDF's) of the number of cycles before failure. In this section we show how to predict these PDF's at any wanted value of the amplitude of forcing from measurements done at another value of the forcing. A summary and discussion are offered in Sect.~\ref{summary}.

\section{ System preparation and protocols} 
\label{simulations}
\subsection{System}
In both protocols of cyclic loads we employ the same glass former, which is composed of a ternary mixture of point particles A, B and C with a concentration ratio A:B:C =  54:29:17, which is embedded in two dimensions. The reason for choosing a ternary rather than binary mixture is to allow a deep quench using Swap Monte Carlo, and see \cite{20POB} for details. The particles interact via a modified Lennard-Jones potential-
\begin{eqnarray}
&&	V_{\alpha,\beta}(r) = 4\epsilon_{\alpha\beta} \Big[\left(\frac{\sigma_{\alpha\beta}}{r}\right)^{12} - \left(\frac{\sigma_{\alpha\beta}}{r}\right)^{6}
	+C_0 \nonumber+ C_2\left(\frac{r}{\sigma_{\alpha\beta}}\right)^{2} \nonumber\\&&+  C_4\left(\frac{r}{\sigma_{\alpha\beta}}\right)^{4}\Big] \ , 
	\label{LJ}
\end{eqnarray}
where $\alpha$ and $\beta$ stand for different types of particle. The potential vanishes at $r_c = 1.75\sigma_{\alpha,\beta}$. This value of $r_c$ is chosen to make the material neither too ductile nor too brittle \cite{11DKPZ}. The constants $C_0, C_2$ and $C_4$ are chosen such that at $V_{\alpha,\beta}(r_c)=V'_{\alpha,\beta}(r_c)=V''_{\alpha,\beta}(r_c)=0$. The energy scales are $\epsilon_{AB}$ = $1.5\epsilon_{AA}$, $\epsilon_{BB}$ = $0.5\epsilon_{AA}$, $\epsilon_{AC}$ = $0.5(\epsilon_{AA} + \epsilon_{AB})$, $\epsilon_{BC}$ = $0.5(\epsilon_{AB} + \epsilon_{BB})$ and $\epsilon_{CC}$ = $0.5(\epsilon_{AA} + \epsilon_{BB})$, with $\epsilon_{AA}$ equal 1.
The ranges of interaction are $\sigma_{AB}$ = $0.8\sigma_{AA}$, $\sigma_{BB}$ = $0.88\sigma_{AA}$, $\sigma_{AC}$ = $0.5(\sigma_{AA} + \sigma_{AB})$, $\sigma_{BC}$ = $0.5(\sigma_{AB} + \sigma_{BB})$ and $\sigma_{CC}$ = $0.5(\sigma_{AA} + \sigma_{BB})$, with $\sigma_{AA}$=1. The mass $m$ of each particle is unity, and the unit of time is $\sqrt{m\sigma_{AA}^2/\epsilon_{AA}}$. Boltzmann's constant is taken as unity. 

To stress the generality and robustness of our scaling approach we have used different preparation protocols. We always begin by equilibrating the ternary mixture in a rectangular box with periodic boundary conditions at a high temperature $T=1$. Subsequently we cool the system in small steps $\Delta T$, and after each step equilibrate the system in NVT conditions. Once we reach $T=0$, the system is equilibrated again with NPT conditions, choosing $P=0$. Once the pressure vanishes, we can remove the periodic boundary conditions and get the system ready for cyclic loading. To avoid spurious periodic cycles at $T=0$  \cite{13RLR} we warm the system up to very low desired temperature, between $T=0.01$ to $T=0.03$. A different protocol prepares the system at $T=0.01$ using Swap Monte Carlo \cite{01GTP}, in which $10^4$ swap steps are performed with
a swap probability $p=0.2$.  The actual conditions for the results displayed below will be presented when needed. 

\subsection{Bending protocol}

\begin{figure}
	\includegraphics[scale=0.45]{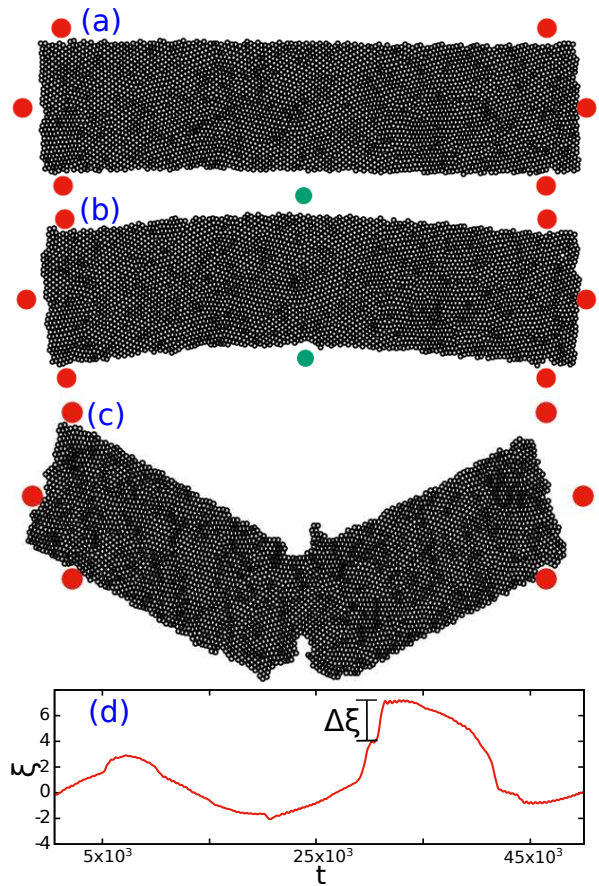}
\caption{Panel (a): the strip of system A with the stoppers and pusher. In red dots we denote the stoppers
that cannot be moved. The green dot is the pusher on which the force $F_{\rm app}$ is applied. Panel (b): the bent strip. Panel (c): The strip being broken after $n_f$ cycles. 
The breaking of the strip is determined in the simulation by observing the disappearance of
curvature in the strip boundary and in the center of mass exceeding $L/2$.
Panel (d): A typical trajectory of the center a mass of the strip.}
\label{diagram}
\end{figure}

 A typical simulation employs
a strip whose length (in the $x$ direction) and width (in the $y$ direction) are $L$ and $W$ respectively. To bend the system in a cyclic fashion we use ``pushers" and ``stoppers", motivated by the experiment of Bonn et al.~\cite{98BKPBM}. The geometry and the placement of ``pushers" and "stoppers" are shown in Fig.~\ref{diagram}. The interaction of both pushers and stoppers with the system particles is the same Lennard-Jones law Eq.~(\ref{LJ}). 
In equilibrium the center of mass is at the origin. To bend the strip in the $\hat{y}$ and $-\hat{y}$ direction we apply a force $F$ at $(0, -W)$ and $(0, W)$ respectively. To prevent  a translational motion of the strip we use four stoppers which have positions pinned at $(-L/2 + 3\sigma_{st}, W/2 + \sigma_{st})$, $(L/2 - 3\sigma_{st}, W/2 + \sigma_{st})$, $(L/2 - 3\sigma_{st}, -W/2 - \sigma_{st})$, $(-L/2 + 3\sigma_{st}, -W/2 - \sigma_{st})$, where $\sigma_{st}$ is the diameter of the stopper. We chose $\sigma_{st} = 5\sigma_{AA}$. A pusher particle with diameter $\sigma_{push} = 4\sigma_{AA}$ is used to apply the external force. These sizes are a multiple of the particle sizes to avoid penetration, and the pusher is smaller than the stoppers to concentrate the force at the center of the strip. To bend the strip in $\hat{y}$ direction we place the pusher at $(0, -W/2)$ and move it in $\hat{y}$ direction by increasing an external force $\B F(t) = F_{\rm app}\sin(\alpha t) \hat{y}$, where $F_{\rm app}$ is the maximum value of applied force. We choose $\alpha = (\frac{\pi}{2})\frac{10^{-3}}{F_{\rm app}}$, such that after a time $t = 10^{3}$ the force increases by one unit. The pusher attains the maximum value $F_{\rm app}$ after traversing a path of length $\xi_{\rm max}$. Then the external force is reduced back to 0 so the the strip can relax and return to its initial state. After that the pusher is moved to $(0, W/2)$ and we apply an external force in the $-\hat{y}$ direction. Now $F(t) = - F_{\rm app}\sin(\alpha t) \hat{y}$, and the pusher reaches its utmost negative position after traversing a length $\xi_{\rm min}$. Finally the force is increased again until it vanishes: one cycle is then completed. These cycles are repeated $n$ times until the strip fails and breaks at some value $n=n_f$.
This bending protocol was applied to three different systems: system A has $N=4228$, $L=130$ and $W=24.4$. System B has $N = 3200$, $L=112$ and $W=21.0$. System C is the largest, with $N=6800$, $L=160$ and $W=30$. The bending protocol was applied to all these systems at $T = 0.01,0.02$ and 0.03. A fourth protocol of preparation employs Swap Monte Carlo at $T=0.01$. Note that the aspect ratio was kept constant at ($L/W \sim 5.33$). For system A we used seven different values of applied external force, $F_{app}$ = 6.25, 6.5, 6.75, 7.0, 7.5, 8.0 and 8.5. For system B we used
the seven values $F_{app}$ = 5.75, 6.0, 6.25, 6.5, 7.0, 7.25, 7.5. For system C the forces were $F_{app}$ =7.5, 7.75, 8, 8.5, 9 and 10.

\subsection{Tensile protocol}

\begin{figure}
	\includegraphics[scale=0.58]{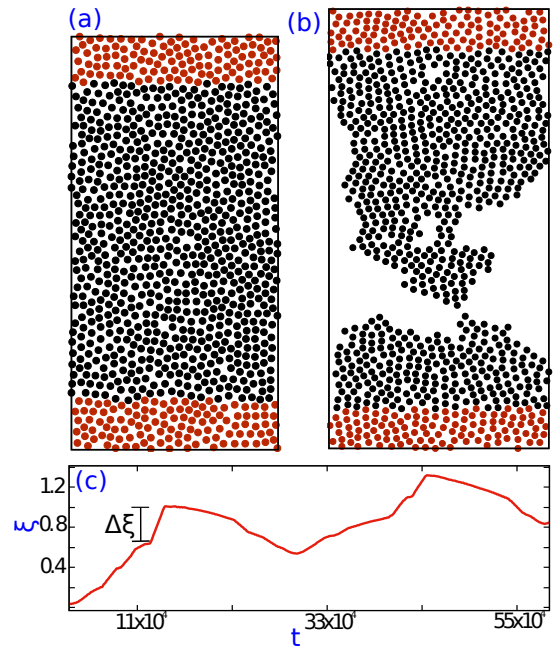}
	\caption{Panel (a): A typical sample of system E. Panel (b): Configuration after the mechanical failure. Panel (c):  A typical trajectory of the center a mass of the strip. The size of the one of the jumps is shown by black vertical line.}
	\label{strip}
	\end{figure}

Tensile loading is a well known method to study the strength of the materials\cite{20PDHK,11DKPZ}. For the tensile periodic protocol we employ two systems, referred to as system E and F.
System E has $N=1000$ and length $L=42$ and width $W=21$. System F contains $N=2000$ particles with  $L=60$ and width $W=30$. The center of mass of the strip is kept at $(W/2, L/2)$ with the coordinates of the four corners being $(0,0)$, $(W,0)$  $(W, L)$  $(0, L)$; see Fig.~\ref{strip} (a) in which system E is shown. Some of the particles of the strip are chosen to make two boundary walls of thickness $5\sigma_{AA}$ at $y = 0$ and $y = L$ (red particles in the figure). To apply a cyclic force in the $\hat{y}$ direction, all the wall particles of the bottom wall are pinned at their position and an equal amount of external force is applied on every particle of the top wall. The force applied on the top wall evolves with time as $F(t) = F_{app}\sin(\alpha t) \hat{y}$. Where $F_{app}$ is the net maximum value of applied force. We choose $\alpha = (\frac{\pi}{2})\frac{10^{-4}}{F_{app}}$ which means the external force takes a time $t = 10^{4}$ to reach from 0 to unity. When the force attains the maximum value $F_{app}$, the external force on the wall is reduced similarly to 0, so the strip can relax to its initial state. Then one cycle is completed. Note that $F_{app}$ is the net applied force on the entire wall; the force on every wall particle is $F_{app}/N_w$, where $N_w$ is the number of particle in the top wall.
The list of applied forces were $F_{app} = 14.25, 14.5, 15.0, 15.5, 16.0, 16.5, 17.0$ for system E and  $F_{app} = 20.0, 20.5, 21.0, 21.5, 22.0$ and $22.5$ for system F.

\section{Definition of measured quantities}  
\label{damage}

Of primary interest in such fatigue experiments is the number of cycles $n_f$ at which the system fails. ``Failure" here is defined as the appearance of macroscopic break in the strip that is not healed by further bending or tensile cycles, see Fig.~\ref{diagram} panel (c) and Fig.~\ref{strip} panel (b).  As expected, $n_f$ is a stochastic variable that is widely distributed, with very large sample to sample fluctuations even for one chosen value of $F_{\rm app}$. It therefore makes sense to focus on the distribution function $P(n_f;F_{\rm app})$ and its mean value, denoted as $\langle n_f \rangle(F_{\rm app}) $; both these quantities depend on $F_{\rm app}$, cf. Sect.~\ref{statistics} below. Two other quantities of interest are (i) the value of $F_{\rm app}$ that results in system's failure in {\em one} cycle. This value is denoted below as $F_Y$ and it depends (for a given system size) on the temperature $T$; and (ii) the value of $F_{\rm app}$ below which we find no failure on the time scale of our simulations \cite{19Gon}. The W\"ohler plots presented below pertain to the range $F_L < F_{\rm app} < F_Y$. We note that when $F_{app}$ is too close to $F_L$, some of the  realizations do not suffer any mechanical damage; they do not fail within the time scale of our simulation. Such realizations are not included in our study. 

The other quantity of major interest is the ``damage" $D_n^s$ that accumulates in every cycle. The definition of damage is not a-priori obvious \cite{08KCAH,08VAH}. We propose here to define the damage as
the energy ``wasted" during plastic events. To introduce this quantity one observes the trajectory of the strip's center of mass, cf Fig.~\ref{diagram} panel (d) and Fig.~\ref{strip} panel (c). The trajectory is smooth modulo temperature fluctuations, but every now and then it suffers a discontinuity when a plastic event is taking place. The trajectory ``jumps" an amount $\Delta \xi$ within
and interval of time $\Delta t$ with 
the force $F(t)$ being fixed. To distinguish from temperature fluctuations we have employed a threshold of $\Delta \xi/\Delta t \ge \epsilon$, with $\epsilon=0.04$ and 0.005 for the bending and tensile cases respectively. The reason for these thresholds is that for lower values of $\epsilon$ the distinction between temperature fluctuations and these events becomes inseparable. These jumps were identified as the damage $F(t) \Delta \xi$ where the value of $F(t)$ was taken from the middle
of the interval $\Delta t$. The damage was added up for all the jumps occurring during
a given $n$th cycle, giving rise to the quantity $D^s_n$,
\begin{equation}
	D_n^s \equiv \sum_k F_k(t) (\Delta \xi)_k
\end{equation}
where the sum on $k$ runs on all the jumps taking place in the $n$th cycle. Finally we are interested in $D_{\rm acc}$ which is defined as the total accumulated damage
during all the cycles until collapse,
\begin{equation}
D_{\rm acc} (n_f) \equiv \sum_{n=1}^{n_f} 	D_n^s \ .
\label{defDacc}
\end{equation}  
This quantity will turn out to be crucial for the understanding of the collapse due to accumulated damage. We note here that small plastic events cannot be safely separated from temperature fluctuations, and therefore our measurements of the damage should be taken as a lower bound on the actual amount of energy spent on plastic events. 

\begin{figure}
\includegraphics[scale=0.60]{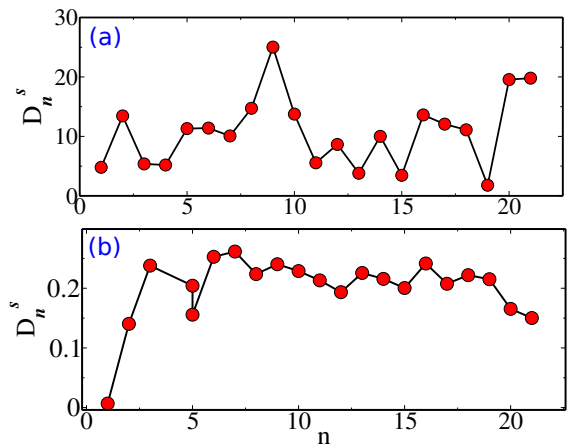}	
	\caption{Panel (a) the damage per cycle $D_n^s$ for a typical sample undergoing
		cyclic bending at $T=0.02$, $F_{\rm app}=7.5$. Panel (b) The damage per cycle 
		for the cyclic tensile protocol. System E, $T=0.02$,  $F_{\rm app}=15$.}
	\label{damagepc}
\end{figure}

\begin{figure}
	\includegraphics[scale=0.57]{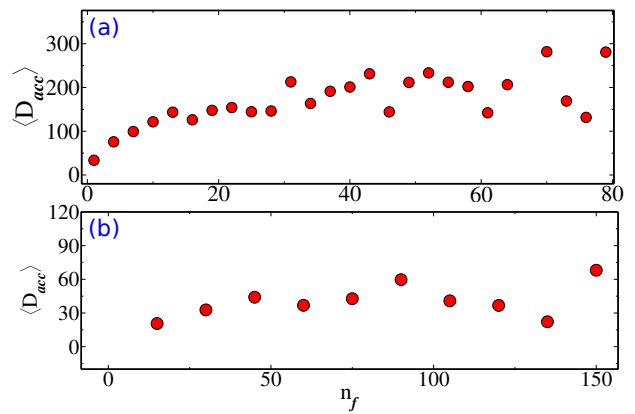}	
	\caption{Panel (a) the accumulated damage $D_{\rm acc} (n_f)$ as a function of $n_f$ for a typical sample of system A undergoing
		cyclic bending at $T=0.02$, $F_{\rm app}=7.5$. Panel (b) The  accumulated damage
		for the cyclic tensile protocol. System E, $T=0.02$,  $F_{\rm app}=15$.}
	\label{damacc}
\end{figure}

Examples of the damage per cycle $D_n^s$ as measured in our protocols are shown in Fig.~\ref{damagepc} for the bending and tensile protocols. We note that the actual values of $D_n^s$ are random and do not depend on the history of the forcing. Both samples
broke after 21 cycles. Examples of the accumulation of damage as a function of $n_f$ are shown in Fig.~\ref{damacc}
for both cyclic bending and tensile protocols.  
It is important to notice that the accumulated damage reaches an asymptotic value which does not depend on $n_f$.

\section{W\"ohler plots and scaling exponents} 
\label{exponents}

As mentioned above, the most typical measurement in engineering contexts is provided by the W\"ohler diagram, relating the number of cycles to failure to the stress level. We therefore present first
the data obtained for both bending and tensile protocols.

\subsection{W\"ohler plots for bending}

\begin{figure*}
	\includegraphics[scale=0.58]{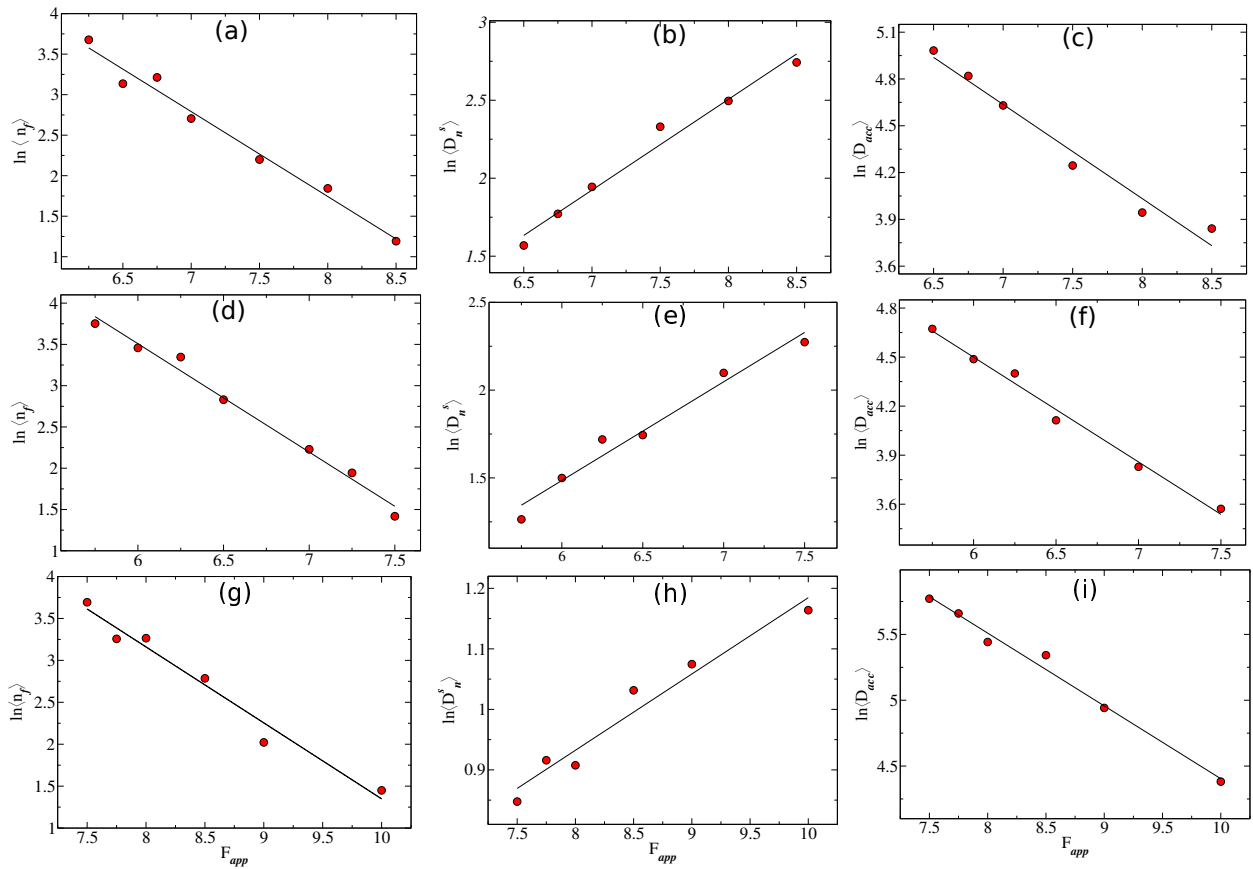}
	\caption{Panels (a), (d) and (g):  W\"ohler plot for systems A, B and C respectively. Panels (b), (e) and (h): average damage per cycle for systems A, B and C respectively. Panels (c), (f) and (i): average accumulated damage for systems A, B and C respectively. }
	\label{wohler}
\end{figure*}

In Fig.~\ref{wohler} we present the average number of cycles to failure $\langle n_f\rangle$ as a function of $F_{\rm app}$ in a log-linear plot for systems A, B and C. The average was computed from about 500 realizations for every value of $F_{\rm app}$.

The data support an exponential dependence of the form
\begin{equation}
	\ln \langle n_f\rangle \approx a (F_Y - F_{\rm app}) \ ,
	\label{nf}
\end{equation}
where $F_Y$ is the applied force that breaks the system in one cycle, and the coefficient $a$ is dimensional, with units of inverse force, to be identified below. For systems A, B and C $aF_Y$ is $10.13\pm 0.48$  and $11.38\pm 0.45$ $10.41\pm 0.66$ respectively. The numerical values of $a$ are  $a\approx 1.05\pm 0.07$ $a\approx 1.31 \pm 0.07$ and $a=0.91\pm 0.06$ respectively. For future reference the parameter of this and further numerical fits are collected in Table~\ref{table}. We note that the range of $F_{\rm app}$ is limited, but this is typical for W\"ohler plots also in engineering experiments, cf. \cite{13JMLC,16PBP,16VEHLZZ}.

\subsection{Damage measures}

A first clue to the origin of Eq.~(\ref{nf}) is provided by the average of the damage per cycle
$\langle D^s_n\rangle$ and its dependence on $F_{\rm app}$, as shown in Fig.~\ref{wohler}.
This is also an exponential, {\em growing} with $ F_{\rm app}$;
\begin{equation}
	\ln \langle D^s_n\rangle \approx D_0+b F_{\rm app} \ ,
	\label{Dsexo}
\end{equation}
where $D_0=-2.13\pm 0.40$, -1.89 $\pm 0.33$ and $-2.4 \pm 0.27$;  $b=0.58\pm 0.05$, 0.56$\pm 0.05$ and $b=0.35\pm 0.03 $ for systems A, B and C respectively. Here again $b$ is a constant with dimension of inverse force.
\begin{table}[h!]
	\begin{center}
		\begin{tabular}{c|c|c|c}
			~&	a & b & c \\
			\hline
			System A & $1.05\pm 0.07$ & $0.58\pm 0.05$ & $0.6\pm 0.04$  \\ \hline
			System B & $1.31\pm 0.07$ & $0.56\pm 0.05$ &$ 0.64\pm 0.03$  \\ \hline
			System C & $0.91\pm 0.08$ & $0.35\pm 0.03$ & $0.55 \pm 0.03$ \\ \hline
		\end{tabular}
	\end{center}
	\caption{The parameters of dimension of inverse force that determine the exponential dependence on the applied force, see text for details.}
	\label{table}
\end{table}
\begin{figure*}
	\includegraphics[scale=0.58]{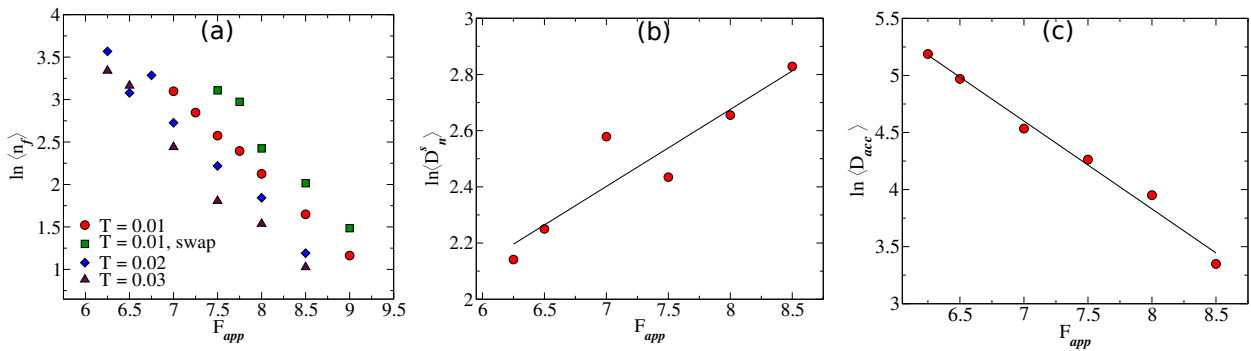}
	\caption{Panel (a): W\"ohler plots for system A at four different conditions, $T=0.01$ (red circles), $T=0.02$ (blue diamonds), $T=0.03$ (maroon triangles) and $T=0.01$ with Swap Monte Carlo. The slope
		$a$ is invariant, being 1.05 within the error bars for all the conditions. Panel (b): average damage per cycle for $T=0.03$ Panel (c): average accumulated damage for $T=0.03$.}
	\label{more}
\end{figure*}

Probably the most interesting numerical finding has to do with $D_{\rm acc}$. We find that with the exception of fragile configurations that break very quickly, for all the strips that survive more than about 20 cycles, $D_{\rm acc}$ attains a constant value that depends on $F_{\rm app}$ but not on the number of cycles. This result will have significant implications as it underlines the fact that accumulated damage is the critical physical quantity that leads to mechanical collapse. The dependence of the average of this quantity on $F_{\rm app}$ is shown in Fig.~\ref{wohler}. Computing the average $\langle D_{\rm acc} \rangle$ the data support again an exponential fit of the form
\begin{equation}
\ln \langle D_{\rm acc} \rangle \approx C  -c F_{\rm app}  \ , 
\label{Daccexp}
\end{equation}
Here $C = 8.86\pm 0.38$, 8.34$\pm 0.23$, $9.93\pm 0.27$ and  $c=0.6\pm 0.04$, 0.64$\pm 0.03$ and $0.55\pm 0.03$ respectively for systems A, B and C. The constant $c$ is the last  parameter with dimension of inverse force. 

In order to demonstrate the generality of robustness of the scaling theory below, we show 
W\"ohler plots for additional bending protocols. In Fig.~\ref{more} we present such plots 
in panel (a). Having four different protocols, with three different temperatures and one having Swap Monte Carlo preparation, we find the the exponent $a$ remains invariant, $a\approx 1.05$ within the error bars. 
In panels (b) and (c) we present the damage analysis for system A at $T=0.03$. The measured values
of the the parameters are $b=0.27\pm 0.06$ and $c=0.77\pm 0.05$.

\subsection{Tensile protocol}

\begin{figure*}
	\includegraphics[scale=0.60]{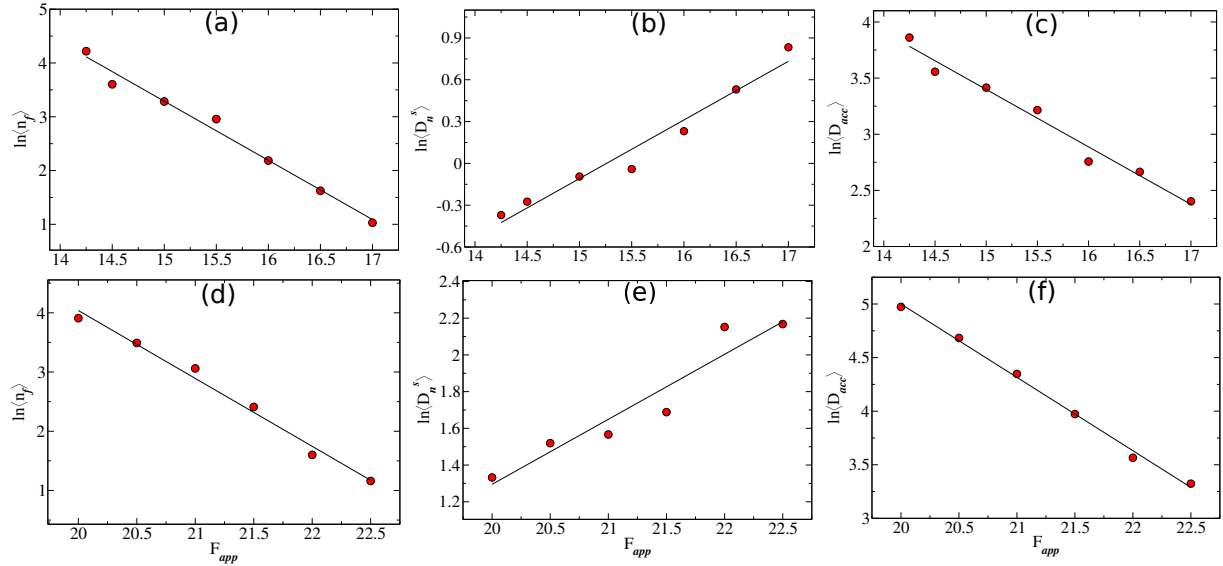}
	\caption{Panels (a) and (d):  W\"ohler plot for systems E and F respectively. Panels (b) and (e): average damage per cycle for systems E and F respectively. Panels (c) and (f): average accumulated damage for systems E and F respectively. }
	\label{tensile}
	\end{figure*}

The analysis of the system response and failure under the tensile protocol follows verbatim
the previous analysis of the bending protocol. In Fig.~\ref{tensile} we present the W\"ohler plots (panels (a) and (d)), the average damage per cycle (panels (b) and (e)) and average accumulated damage (panels (c) and (f)) for systems E and F for T=0.02. We find again similar scaling laws Eqs.~(\ref{nf}), (\ref{Dsexo})
and (\ref{Daccexp}), with the scaling exponents as collected in table \ref{tens}. The pre-exponential constants are $aF_{\rm y}$ is $19.79\pm 0.95$  and $26.97\pm 1.39$, $D_0=-7.0\pm 0.56$ and $-5.79\pm 1.16$ and $C=11.07\pm 0.56$ and $18.69\pm 0.45$ for systems E and F respectively. 

\begin{table}[h!]
	\begin{center}
		\begin{tabular}{c|c|c|c}
			~&	a & b & c \\
			\hline
			System E & $1.10\pm 0.06$ & $0.46\pm 0.04$ & $0.52\pm 0.04$  \\ \hline
			System F & $1.15\pm 0.07$ & $0.36\pm 0.05$ &$ 0.69\pm 0.02$  \\ \hline
		\end{tabular}
	\end{center}
	\caption{The parameters of dimension of inverse force that determine the exponential dependence on the applied force for the tensile protocol.}
	\label{tens}
\end{table}

\section{Scaling theory} 
\label{scaling}

At first sight the results of the numerical simulations appear confusing, with a variety of numerical values of the parameters $a$, $b$ and $c$ as summarized in tables \ref{table} and \ref{tens}.
In this section we offer a scaling theory to rationalize the numerical values of these coefficients. We will make the point that the proposed theory is a first step in the solution of the long standing riddle of W\"ohler diagrams.

The first observation is that Eqs.~(\ref{Dsexo}) and (\ref{Daccexp}) provide
a physical reason for the W\"ohler relation Eq.~(\ref{nf}). The idea is that the average number
of cycles to failure will be determined by the following ratio:
\begin{equation}
	\langle n_f \rangle \sim  \langle D_{\rm acc}\rangle /	\langle D_n^s\rangle \ ,
	\label{explain}	
\end{equation}
up to a constant of the order of unity. Simply the amount of damage per cycle accumulates to the (approximately) constant
value $\langle D_{\rm acc}\rangle$ during $\langle n_f\rangle $ cycles. We realize that scaling theory can only provide predictions up to constants of the order of unity, but nevertheless it is worthwhile to see how well we can explain the numerical results. Plugging in Eq.~(\ref{explain}) the
numerical values of the pre-exponential constants and the values of $b$ and $c$ for both systems we estimate
\begin{eqnarray}
\ln	\langle n_f \rangle &\approx& 10.99\pm 0.78 - (1.18\pm 0.09) F_{\rm app}~\text{system A} \nonumber \ ,
\\
\ln	\langle n_f \rangle &\approx& 10.23\pm 0.56 - (1.2\pm 0.08) F_{\rm app}~\text{system B} \ , \nonumber\\
\ln	\langle n_f \rangle &\approx& 12.33\pm 0.54 - (0.9\pm 0.06) F_{\rm app}~\text{system C} \ , \nonumber\\
\ln	\langle n_f \rangle &\approx& 17.51\pm 0.56 - (0.98\pm 0.08) F_{\rm app}~\text{system E} \ , \nonumber \\
\ln	\langle n_f \rangle &\approx& 24.87\pm 1.78 - (1.05\pm 0.07) F_{\rm app}~\text{system F}  \ .
\label{comparison}
\end{eqnarray} 
These result are consistent within the error bars with the numerical  simulations Eq.~(\ref{nf}), for both bending and tensile protocols.  We therefore can propose a scaling relation: 
\begin{equation}
	a=b+c\ .
	\label{wow}
\end{equation}

We reiterate at this point that all these three numbers ($a, b$ and $c$) are dimensional, being inverse forces. Until now we do not have any prior knowledge of these dimensional coefficients. To seek this information we examine again the values of $\langle D_{\rm acc} \rangle$ and realize that they span one order of $e$. We will therefore construct an approximate scale
of damage by averaging $\langle D_{\rm acc} \rangle$ over its range $F_L\le F_{\rm app} \le F_Y$, creating an average of averages, denoted as 
$\overline{\langle {D_{\rm acc}} \rangle}$. Our simulations suggest that $F_L = 5.0, 4.25, 6.50, 13.0, 18.50$ and $F_Y = 9.97, 8.85, 11.76, 18.83, 24.85$ for system A, B, C, E and F respectively. From the numerics we measure $\overline{\langle {D_{\rm acc}} \rangle}\approx 111$ for system A , 88 for system B, 188 for system C, 56 for system E and 84 for system F.

With this scale in mind, consider the parameter $b$ in Eq.~(\ref{Dsexo}). It has the dimension of inverse force and must be independent of $F_{\rm app}$. The only energy scale available that is independent of $F_{\rm app}$ is  $\overline{\langle {D_{\rm acc}} \rangle}$, and the length scale associated with damage must be $\ell_D\equiv \sqrt{LW}$, since plastic events can appear anywhere in the area of the strip. For system A $\ell_D=56.36$ , for system B $\ell_D=48.5$ and for system C $\ell_D=69.3$. In the case of tensile forcing $\ell_D\equiv \sqrt{L_{a}W}$, where $L_a$ is the length of the strip without the walls. So $L_a = L - 2L_{wall}$, where $L_{wall}$ is the width of the walls which is approximately $5\sigma_{AA}$. We find that for system E and F $\ell_D$ is 25.92 and 38.73. We thus estimate
\begin{eqnarray}
b\approx \ell_D/\overline{\langle D_{\rm acc} \rangle} &\approx& 0.51 \quad \text{system A} \ ,\\
&\approx& 0.55 \quad \text{system B} \ , \\
&\approx& 0.37 \quad \text{system C} \ , \\
&\approx& 0.46 \quad \text{system E} \ ,\\
&\approx& 0.46 \quad \text{system F} \ .
\label{amazing}	
\end{eqnarray}
Taking into account the approximate nature of the scale $\overline{\langle {D_{\rm acc}} \rangle}$ and the fact that it is a lower bound, we consider the results to be in good agreement with the data. 

Regarding the numerical value of $a$, in the case of bending we expect that failure starts with a micro-crack at the upper {\em or} lower boundary, so the relevant scale for Eq.~(\ref{nf}) is $L$. For tensile forcing, since the micro crack can appear in any side of the strip, the relevant scale is $2L_a$. Therefore we estimate
\begin{eqnarray}
a\approx L/\overline{\langle D_{\rm acc} \rangle} &\approx& 1.17 \quad \text{system A} \ ,\\
&\approx& 1.27 \quad \text{system B} \ , \\
&\approx& 0.85 \quad \text{system C} \ , \\
a\approx 2L_a/\overline{\langle D_{\rm acc} \rangle} &\approx& 1.14 \quad \text{system E} \,  \\
&\approx& 1.19 \quad \text{system F} \ .
\label{nice}	
\end{eqnarray}

We note that the resulting estimates appear very close to reality (withing the error bars) but they do change wildly between different system sizes and protocols. Therefore one should ask whether the present approach can provide predictability for {\em a given system} 
at different values of $F_{\rm app}$. In the next section we answer this question in the affirmative using probability distribution functions.

\section{Probability Distribution Functions}
\label{statistics}
\begin{figure*}
	\includegraphics[scale=0.62]{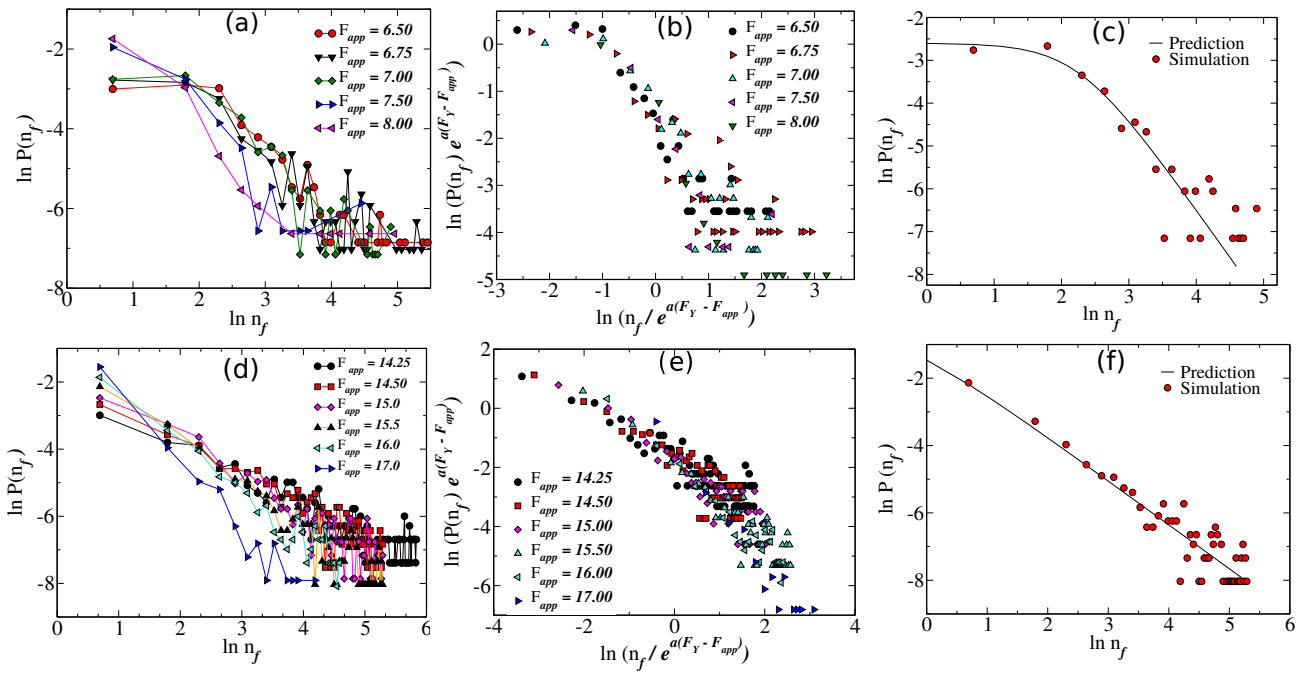}
	\caption{Panel (a) and (d): Probability distribution ($P(n_f)$)of number of cycle to failure ($n_f$) for system system A and F. Panel (b) and (e): Scaling of $P(n_f)$ using the scaling relation. Panel (c) and (f): Prediction of $P(n_f)$ from the scaling function $g(x)$ for system A and E for $F_{app} = 7.0$ and $F_{app} = 16.0$ respectively.}
	\label{pdfs}
\end{figure*}
\subsection{Rescaling}
Besides averages, the simulation provide us also with the probability distribution functions (PDF) of all the measured quantities. In this section we focus on the most important PDF of the number of cycles for failure $n_f$. In Fig.~\ref{pdfs} panels (a) and (d) we present the PDF's for system A and system E respectively. We observe that for different values of $F_{\rm app}$ these PDF's spread over three orders of magnitude (in $e$) in terms of $n_f$. We can therefore use our scaling theory
to collapse these PDF's in a way that will allow us eventually to predict the PDf at one value of $F_{\rm app}$ from another value. 

To collapse the date we re-plot the PDF's as scaling functions in the form
\begin{equation}
P(n_f; \langle n_f\rangle) = \langle n_f\rangle ^{-1} g(n_f/\langle n_f\rangle)	\ . 
\end{equation}
Moreover, we will use the scaling law Eq.~(\ref{nf}) in the form 
\begin{equation}
	\langle n_f\rangle \approx e^{a(F_Y -F_{\rm app})}  \ .
	\label{rescale}
\end{equation}
The result of this rescaling is presented in Fig.~\ref{pdfs} in panels (b) and (e) for systems A and E respectively. While not a perfect collapse, the PDF's are now all mixed together with the spread being attributed to the noisy tail of the un-rescaled PDF's. We presently find
``best fits" to the scaling function $g(x)$ in the form 
\begin{equation}
g(x) =\frac{\alpha} {1+\gamma x^{\beta}}	 \ ,
\label{fit}
	\end{equation}
where $\alpha, \beta, \gamma$ are fit parameters. 

\subsection{Predictability} 

The utility of Eqs.~(\ref{rescale}) and (\ref{fit}) is in allowing us to predict the PDF for $n_f$ at a chosen value of $F_{\rm app}$ from measuring the PDF at another value of $F_{\rm app}$. Examples of such
prediction are offered in panels (c) and (f) for bending and tensile
protocols respectively. In panel (c) the fit for the collapse of the PDF in panel (b) is used to predict the PDF at $F_{\rm app}=7$ for system A. The prediction is shown in continuous line, and it should be compared with the data obtained at  $F_{\rm app}=7$ which is presented in dots. In panel (f) the same is done, using the collapsed PDF from panel (e) to predict the PDF for system E for $F_{\rm app}=16$. We stress that these are typical results, they were not selected for their quality. We trust that this kind of predictability can be put to advantageous use in mechanical engineering.

\section{Summary and discussion}
\label{summary}
In summary, we offer a scaling theory of W\"ohler plots, based on the idea that accumulated damage is the fundamental cause for failure, joined with the discovery that this quantity appears constant for system failing at different values of $\langle n_f \rangle$. This phenomenological finding indicates that memory plays a crucial role in fatigue; one can fail after many cycle if the damage accrued in each cycle is small, or in a few cycles if the damage in each happens to be large. What matters is the accumulated damage that seems to have a (presumably material and temperature dependent) limit \cite{07HDJS}. The measurement of this quantity is not trivial due to the difficulty of distinguishing small plastic events from temperature fluctuations. Focusing on this quantity allows understanding of the exponential dependence of the average number of cycles to failure on the applied force. Moreover, introducing the natural scales for failure on the boundary and plastic events in the bulk, the numerical values of the parameters $a$, $b$ and $c$ could be rationalized. We reiterate that we could not discern any temperature dependence in our exponents, excluding any Arrhenius type mechanism of barrier crossing.

It would be very useful to test the ideas presented in this Letter in experiments. The experimenters will need to come up with a robust method to estimate the damage done in each cycle and its accumulated counterpart. If it turned out that the accumulated damage is indeed  independent of the average number of cycles for failure, the path for a scaling theory of the type presented here would open. In parallel, in future work the present effort would continue using numerical simulations and additional theoretical developments.  

\acknowledgments This work has been supported in part by the Minerva Foundation, Munich, Germany, through the Minerva Center for Aging at the Weizmann Institute of Science.

\bibliography{ALL}

\begin{thebibliography}{22}%
\makeatletter
\providecommand \@ifxundefined [1]{%
 \@ifx{#1\undefined}
}%
\providecommand \@ifnum [1]{%
 \ifnum #1\expandafter \@firstoftwo
 \else \expandafter \@secondoftwo
 \fi
}%
\providecommand \@ifx [1]{%
 \ifx #1\expandafter \@firstoftwo
 \else \expandafter \@secondoftwo
 \fi
}%
\providecommand \natexlab [1]{#1}%
\providecommand \enquote  [1]{``#1''}%
\providecommand \bibnamefont  [1]{#1}%
\providecommand \bibfnamefont [1]{#1}%
\providecommand \citenamefont [1]{#1}%
\providecommand \href@noop [0]{\@secondoftwo}%
\providecommand \href [0]{\begingroup \@sanitize@url \@href}%
\providecommand \@href[1]{\@@startlink{#1}\@@href}%
\providecommand \@@href[1]{\endgroup#1\@@endlink}%
\providecommand \@sanitize@url [0]{\catcode `\\12\catcode `\$12\catcode
  `\&12\catcode `\#12\catcode `\^12\catcode `\_12\catcode `\%12\relax}%
\providecommand \@@startlink[1]{}%
\providecommand \@@endlink[0]{}%
\providecommand \url  [0]{\begingroup\@sanitize@url \@url }%
\providecommand \@url [1]{\endgroup\@href {#1}{\urlprefix }}%
\providecommand \urlprefix  [0]{URL }%
\providecommand \Eprint [0]{\href }%
\providecommand \doibase [0]{http://dx.doi.org/}%
\providecommand \selectlanguage [0]{\@gobble}%
\providecommand \bibinfo  [0]{\@secondoftwo}%
\providecommand \bibfield  [0]{\@secondoftwo}%
\providecommand \translation [1]{[#1]}%
\providecommand \BibitemOpen [0]{}%
\providecommand \bibitemStop [0]{}%
\providecommand \bibitemNoStop [0]{.\EOS\space}%
\providecommand \EOS [0]{\spacefactor3000\relax}%
\providecommand \BibitemShut  [1]{\csname bibitem#1\endcsname}%
\let\auto@bib@innerbib\@empty
\bibitem [{\citenamefont {Bhowmik}\ \emph {et~al.}(2021)\citenamefont
  {Bhowmik}, \citenamefont {Hentschel},\ and\ \citenamefont
  {Procaccia}}]{21BHP}%
  \BibitemOpen
  \bibfield  {author} {\bibinfo {author} {\bibfnamefont {B.~P.}\ \bibnamefont
  {Bhowmik}}, \bibinfo {author} {\bibfnamefont {H.~G.~E.}\ \bibnamefont
  {Hentschel}}, \ and\ \bibinfo {author} {\bibfnamefont {I.}~\bibnamefont
  {Procaccia}},\ }\href@noop {} {\enquote {\bibinfo {title} {Fatigue and
  collapse of cyclically bent strip of amorphous solid},}\ } (\bibinfo {year}
  {2021}),\ \Eprint {http://arxiv.org/abs/2103.03040} {arXiv:2103.03040
  [cond-mat.soft]} \BibitemShut {NoStop}%
\bibitem [{\citenamefont {Miles}(1954)}]{54Mil}%
  \BibitemOpen
  \bibfield  {author} {\bibinfo {author} {\bibfnamefont {J.~W.}\ \bibnamefont
  {Miles}},\ }\href@noop {} {\bibfield  {journal} {\bibinfo  {journal} {Journal
  of the Aeronautical Sciences}\ }\textbf {\bibinfo {volume} {21}},\ \bibinfo
  {pages} {753} (\bibinfo {year} {1954})}\BibitemShut {NoStop}%
\bibitem [{\citenamefont {Dowell}(1966)}]{66Dow}%
  \BibitemOpen
  \bibfield  {author} {\bibinfo {author} {\bibfnamefont {E.~H.}\ \bibnamefont
  {Dowell}},\ }\href {\doibase 10.2514/3.3658} {\bibfield  {journal} {\bibinfo
  {journal} {AIAA Journal}\ }\textbf {\bibinfo {volume} {4}},\ \bibinfo {pages}
  {1267} (\bibinfo {year} {1966})}\BibitemShut {NoStop}%
\bibitem [{\citenamefont {Dietmann}\ \emph {et~al.}(1989)\citenamefont
  {Dietmann}, \citenamefont {Bhongbhibhat},\ and\ \citenamefont
  {Schmid}}]{89DBS}%
  \BibitemOpen
  \bibfield  {author} {\bibinfo {author} {\bibfnamefont {H.}~\bibnamefont
  {Dietmann}}, \bibinfo {author} {\bibfnamefont {T.}~\bibnamefont
  {Bhongbhibhat}}, \ and\ \bibinfo {author} {\bibfnamefont {A.}~\bibnamefont
  {Schmid}},\ }in\ \href@noop {} {\emph {\bibinfo {booktitle} {ICBMFF3}}}\
  (\bibinfo {year} {1989})\BibitemShut {NoStop}%
\bibitem [{\citenamefont {Klevtsov}\ and\ \citenamefont {Crane}(1994)}]{94KC}%
  \BibitemOpen
  \bibfield  {author} {\bibinfo {author} {\bibfnamefont {I.}~\bibnamefont
  {Klevtsov}}\ and\ \bibinfo {author} {\bibfnamefont {R.}~\bibnamefont
  {Crane}},\ }\href {\doibase 10.1115/1.2929563} {\bibfield  {journal}
  {\bibinfo  {journal} {Journal of Pressure Vessel Technology}\ }\textbf
  {\bibinfo {volume} {116}},\ \bibinfo {pages} {110} (\bibinfo {year}
  {1994})}\BibitemShut {NoStop}%
\bibitem [{\citenamefont {Freudenthal}\ \emph {et~al.}(1953)\citenamefont
  {Freudenthal}, \citenamefont {Gumbel},\ and\ \citenamefont {Gough}}]{53FGG}%
  \BibitemOpen
  \bibfield  {author} {\bibinfo {author} {\bibfnamefont {A.~M.}\ \bibnamefont
  {Freudenthal}}, \bibinfo {author} {\bibfnamefont {E.~J.}\ \bibnamefont
  {Gumbel}}, \ and\ \bibinfo {author} {\bibfnamefont {H.~J.}\ \bibnamefont
  {Gough}},\ }\href {\doibase 10.1098/rspa.1953.0024} {\bibfield  {journal}
  {\bibinfo  {journal} {Proc. R. Soc. A Math. Phys. Eng. Sci.}\ }\textbf
  {\bibinfo {volume} {216}},\ \bibinfo {pages} {309} (\bibinfo {year}
  {1953})}\BibitemShut {NoStop}%
\bibitem [{\citenamefont {Chang}\ and\ \citenamefont {Chang}(1987)}]{87CC}%
  \BibitemOpen
  \bibfield  {author} {\bibinfo {author} {\bibfnamefont {F.-K.}\ \bibnamefont
  {Chang}}\ and\ \bibinfo {author} {\bibfnamefont {K.-Y.}\ \bibnamefont
  {Chang}},\ }\href {\doibase 10.1177/002199838702100904} {\bibfield  {journal}
  {\bibinfo  {journal} {Journal of Composite Materials}\ }\textbf {\bibinfo
  {volume} {21}},\ \bibinfo {pages} {834} (\bibinfo {year} {1987})}\BibitemShut
  {NoStop}%
\bibitem [{\citenamefont {Lapczyk}\ and\ \citenamefont {Hurtado}(2007)}]{07LH}%
  \BibitemOpen
  \bibfield  {author} {\bibinfo {author} {\bibfnamefont {I.}~\bibnamefont
  {Lapczyk}}\ and\ \bibinfo {author} {\bibfnamefont {J.~A.}\ \bibnamefont
  {Hurtado}},\ }\href {\doibase
  https://doi.org/10.1016/j.compositesa.2007.01.017} {\bibfield  {journal}
  {\bibinfo  {journal} {Appl. Sci. and Manufacturing}\ }\textbf {\bibinfo
  {volume} {38}},\ \bibinfo {pages} {2333} (\bibinfo {year}
  {2007})}\BibitemShut {NoStop}%
\bibitem [{\citenamefont {Gorbatikh}\ and\ \citenamefont {Lomov}(2016)}]{16TV}%
  \BibitemOpen
  \bibfield  {author} {\bibinfo {author} {\bibfnamefont {L.}~\bibnamefont
  {Gorbatikh}}\ and\ \bibinfo {author} {\bibfnamefont {S.}~\bibnamefont
  {Lomov}},\ }in\ \href {\doibase
  https://doi.org/10.1016/B978-1-78242-286-0.00003-0} {\emph {\bibinfo
  {booktitle} {Modeling Damage, Fatigue and Failure of Composite Materials}}},\
  \bibinfo {series and number} {Woodhead Publishing Series in Composites
  Science and Engineering},\ \bibinfo {editor} {edited by\ \bibinfo {editor}
  {\bibfnamefont {R.}~\bibnamefont {Talreja}}\ and\ \bibinfo {editor}
  {\bibfnamefont {J.}~\bibnamefont {Varna}}}\ (\bibinfo  {publisher} {Woodhead
  Publishing},\ \bibinfo {year} {2016})\ pp.\ \bibinfo {pages}
  {41--59}\BibitemShut {NoStop}%
\bibitem [{\citenamefont {Parmar}\ \emph {et~al.}(2020)\citenamefont {Parmar},
  \citenamefont {Ozawa},\ and\ \citenamefont {Berthier}}]{20POB}%
  \BibitemOpen
  \bibfield  {author} {\bibinfo {author} {\bibfnamefont {A.~D.}\ \bibnamefont
  {Parmar}}, \bibinfo {author} {\bibfnamefont {M.}~\bibnamefont {Ozawa}}, \
  and\ \bibinfo {author} {\bibfnamefont {L.}~\bibnamefont {Berthier}},\
  }\href@noop {} {\bibfield  {journal} {\bibinfo  {journal} {Physical Review
  Letters}\ }\textbf {\bibinfo {volume} {125}},\ \bibinfo {pages} {085505}
  (\bibinfo {year} {2020})}\BibitemShut {NoStop}%
\bibitem [{\citenamefont {Dauchot}\ \emph {et~al.}(2011)\citenamefont
  {Dauchot}, \citenamefont {Karmakar}, \citenamefont {Procaccia},\ and\
  \citenamefont {Zylberg}}]{11DKPZ}%
  \BibitemOpen
  \bibfield  {author} {\bibinfo {author} {\bibfnamefont {O.}~\bibnamefont
  {Dauchot}}, \bibinfo {author} {\bibfnamefont {S.}~\bibnamefont {Karmakar}},
  \bibinfo {author} {\bibfnamefont {I.}~\bibnamefont {Procaccia}}, \ and\
  \bibinfo {author} {\bibfnamefont {J.}~\bibnamefont {Zylberg}},\ }\href
  {\doibase 10.1103/PhysRevE.84.046105} {\bibfield  {journal} {\bibinfo
  {journal} {Phys. Rev. E}\ }\textbf {\bibinfo {volume} {84}},\ \bibinfo
  {pages} {046105} (\bibinfo {year} {2011})}\BibitemShut {NoStop}%
\bibitem [{\citenamefont {Regev}\ \emph {et~al.}(2013)\citenamefont {Regev},
  \citenamefont {Lookman},\ and\ \citenamefont {Reichhardt}}]{13RLR}%
  \BibitemOpen
  \bibfield  {author} {\bibinfo {author} {\bibfnamefont {I.}~\bibnamefont
  {Regev}}, \bibinfo {author} {\bibfnamefont {T.}~\bibnamefont {Lookman}}, \
  and\ \bibinfo {author} {\bibfnamefont {C.}~\bibnamefont {Reichhardt}},\
  }\href {\doibase 10.1103/PhysRevE.88.062401} {\bibfield  {journal} {\bibinfo
  {journal} {Phys. Rev. E}\ }\textbf {\bibinfo {volume} {88}},\ \bibinfo
  {pages} {062401} (\bibinfo {year} {2013})}\BibitemShut {NoStop}%
\bibitem [{\citenamefont {Grigera}\ and\ \citenamefont {Parisi}(2001)}]{01GTP}%
  \BibitemOpen
  \bibfield  {author} {\bibinfo {author} {\bibfnamefont {T.~S.}\ \bibnamefont
  {Grigera}}\ and\ \bibinfo {author} {\bibfnamefont {G.}~\bibnamefont
  {Parisi}},\ }\href {\doibase 10.1103/PhysRevE.63.045102} {\bibfield
  {journal} {\bibinfo  {journal} {Phys. Rev. E}\ }\textbf {\bibinfo {volume}
  {63}},\ \bibinfo {pages} {045102} (\bibinfo {year} {2001})}\BibitemShut
  {NoStop}%
\bibitem [{\citenamefont {Bonn}\ \emph {et~al.}(1998)\citenamefont {Bonn},
  \citenamefont {Kellay}, \citenamefont {Prochnow}, \citenamefont
  {Ben-Djemiaa},\ and\ \citenamefont {Meunier}}]{98BKPBM}%
  \BibitemOpen
  \bibfield  {author} {\bibinfo {author} {\bibfnamefont {D.}~\bibnamefont
  {Bonn}}, \bibinfo {author} {\bibfnamefont {H.}~\bibnamefont {Kellay}},
  \bibinfo {author} {\bibfnamefont {M.}~\bibnamefont {Prochnow}}, \bibinfo
  {author} {\bibfnamefont {K.}~\bibnamefont {Ben-Djemiaa}}, \ and\ \bibinfo
  {author} {\bibfnamefont {J.}~\bibnamefont {Meunier}},\ }\href@noop {}
  {\bibfield  {journal} {\bibinfo  {journal} {Science}\ }\textbf {\bibinfo
  {volume} {280}},\ \bibinfo {pages} {265} (\bibinfo {year}
  {1998})}\BibitemShut {NoStop}%
\bibitem [{\citenamefont {Paul}\ \emph {et~al.}(2020)\citenamefont {Paul},
  \citenamefont {Dasgupta}, \citenamefont {Horbach},\ and\ \citenamefont
  {Karmakar}}]{20PDHK}%
  \BibitemOpen
  \bibfield  {author} {\bibinfo {author} {\bibfnamefont {K.}~\bibnamefont
  {Paul}}, \bibinfo {author} {\bibfnamefont {R.}~\bibnamefont {Dasgupta}},
  \bibinfo {author} {\bibfnamefont {J.}~\bibnamefont {Horbach}}, \ and\
  \bibinfo {author} {\bibfnamefont {S.}~\bibnamefont {Karmakar}},\ }\href
  {\doibase 10.1103/PhysRevResearch.2.042012} {\bibfield  {journal} {\bibinfo
  {journal} {Phys. Rev. Research}\ }\textbf {\bibinfo {volume} {2}},\ \bibinfo
  {pages} {042012} (\bibinfo {year} {2020})}\BibitemShut {NoStop}%
\bibitem [{\citenamefont {Gonz{\'a}lez-Vel{\'a}zquez}(2019)}]{19Gon}%
  \BibitemOpen
  \bibfield  {author} {\bibinfo {author} {\bibfnamefont {J.~L.}\ \bibnamefont
  {Gonz{\'a}lez-Vel{\'a}zquez}},\ }\href@noop {} {\emph {\bibinfo {title}
  {Mechanical behavior and fracture of engineering materials}}}\ (\bibinfo
  {publisher} {Springer},\ \bibinfo {year} {2019})\BibitemShut {NoStop}%
\bibitem [{\citenamefont {Kun}\ \emph {et~al.}(2008)\citenamefont {Kun},
  \citenamefont {Carmona}, \citenamefont {Andrade},\ and\ \citenamefont
  {Herrmann}}]{08KCAH}%
  \BibitemOpen
  \bibfield  {author} {\bibinfo {author} {\bibfnamefont {F.}~\bibnamefont
  {Kun}}, \bibinfo {author} {\bibfnamefont {H.~A.}\ \bibnamefont {Carmona}},
  \bibinfo {author} {\bibfnamefont {J.~S.}\ \bibnamefont {Andrade}}, \ and\
  \bibinfo {author} {\bibfnamefont {H.~J.}\ \bibnamefont {Herrmann}},\ }\href
  {\doibase 10.1103/PhysRevLett.100.094301} {\bibfield  {journal} {\bibinfo
  {journal} {Phys. Rev. Lett.}\ }\textbf {\bibinfo {volume} {100}},\ \bibinfo
  {pages} {094301} (\bibinfo {year} {2008})}\BibitemShut {NoStop}%
\bibitem [{\citenamefont {Vieira}\ \emph {et~al.}(2008)\citenamefont {Vieira},
  \citenamefont {Andrade},\ and\ \citenamefont {Herrmann}}]{08VAH}%
  \BibitemOpen
  \bibfield  {author} {\bibinfo {author} {\bibfnamefont {A.~P.}\ \bibnamefont
  {Vieira}}, \bibinfo {author} {\bibfnamefont {J.~S.}\ \bibnamefont {Andrade}},
  \ and\ \bibinfo {author} {\bibfnamefont {H.~J.}\ \bibnamefont {Herrmann}},\
  }\href {\doibase 10.1103/PhysRevLett.100.195503} {\bibfield  {journal}
  {\bibinfo  {journal} {Phys. Rev. Lett.}\ }\textbf {\bibinfo {volume} {100}},\
  \bibinfo {pages} {195503} (\bibinfo {year} {2008})}\BibitemShut {NoStop}%
\bibitem [{\citenamefont {Jegou}\ \emph {et~al.}(2013)\citenamefont {Jegou},
  \citenamefont {Marco}, \citenamefont {{Le Saux}},\ and\ \citenamefont
  {Calloch}}]{13JMLC}%
  \BibitemOpen
  \bibfield  {author} {\bibinfo {author} {\bibfnamefont {L.}~\bibnamefont
  {Jegou}}, \bibinfo {author} {\bibfnamefont {Y.}~\bibnamefont {Marco}},
  \bibinfo {author} {\bibfnamefont {V.}~\bibnamefont {{Le Saux}}}, \ and\
  \bibinfo {author} {\bibfnamefont {S.}~\bibnamefont {Calloch}},\ }\href
  {\doibase https://doi.org/10.1016/j.ijfatigue.2012.09.007} {\bibfield
  {journal} {\bibinfo  {journal} {International Journal of Fatigue}\ }\textbf
  {\bibinfo {volume} {47}},\ \bibinfo {pages} {259} (\bibinfo {year}
  {2013})}\BibitemShut {NoStop}%
\bibitem [{\citenamefont {Pineau}\ \emph {et~al.}(2016)\citenamefont {Pineau},
  \citenamefont {{Amine Benzerga}},\ and\ \citenamefont {Pardoen}}]{16PBP}%
  \BibitemOpen
  \bibfield  {author} {\bibinfo {author} {\bibfnamefont {A.}~\bibnamefont
  {Pineau}}, \bibinfo {author} {\bibfnamefont {A.}~\bibnamefont {{Amine
  Benzerga}}}, \ and\ \bibinfo {author} {\bibfnamefont {T.}~\bibnamefont
  {Pardoen}},\ }\href {\doibase https://doi.org/10.1016/j.actamat.2015.07.049}
  {\bibfield  {journal} {\bibinfo  {journal} {Acta Materialia}\ }\textbf
  {\bibinfo {volume} {107}},\ \bibinfo {pages} {508} (\bibinfo {year}
  {2016})}\BibitemShut {NoStop}%
\bibitem [{\citenamefont {Valiev}\ \emph {et~al.}(2016)\citenamefont {Valiev},
  \citenamefont {Estrin}, \citenamefont {Horita}, \citenamefont {Langdon},
  \citenamefont {Zehetbauer},\ and\ \citenamefont {Zhu}}]{16VEHLZZ}%
  \BibitemOpen
  \bibfield  {author} {\bibinfo {author} {\bibfnamefont {R.}~\bibnamefont
  {Valiev}}, \bibinfo {author} {\bibfnamefont {Y.}~\bibnamefont {Estrin}},
  \bibinfo {author} {\bibfnamefont {Z.}~\bibnamefont {Horita}}, \bibinfo
  {author} {\bibfnamefont {T.}~\bibnamefont {Langdon}}, \bibinfo {author}
  {\bibfnamefont {M.}~\bibnamefont {Zehetbauer}}, \ and\ \bibinfo {author}
  {\bibfnamefont {Y.}~\bibnamefont {Zhu}},\ }\href {\doibase
  10.1080/21663831.2015.1060543} {\bibfield  {journal} {\bibinfo  {journal}
  {Materials Research Letters}\ }\textbf {\bibinfo {volume} {4}},\ \bibinfo
  {pages} {1} (\bibinfo {year} {2016})}\BibitemShut {NoStop}%
\bibitem [{\citenamefont {Harmon}\ \emph {et~al.}(2007)\citenamefont {Harmon},
  \citenamefont {Demetriou}, \citenamefont {Johnson},\ and\ \citenamefont
  {Samwer}}]{07HDJS}%
  \BibitemOpen
  \bibfield  {author} {\bibinfo {author} {\bibfnamefont {J.~S.}\ \bibnamefont
  {Harmon}}, \bibinfo {author} {\bibfnamefont {M.~D.}\ \bibnamefont
  {Demetriou}}, \bibinfo {author} {\bibfnamefont {W.~L.}\ \bibnamefont
  {Johnson}}, \ and\ \bibinfo {author} {\bibfnamefont {K.}~\bibnamefont
  {Samwer}},\ }\href@noop {} {\bibfield  {journal} {\bibinfo  {journal}
  {Physical Review Letters}\ }\textbf {\bibinfo {volume} {99}},\ \bibinfo
  {pages} {135502} (\bibinfo {year} {2007})}\BibitemShut {NoStop}%
\end{thebibliography}%

\end{document}